\documentclass[letterpaper,twocolumn,american,showpacs,prl,aps,superscriptaddress]{revtex4}
\usepackage[latin1]{inputenc}
\usepackage{bm}
\usepackage{multirow}
\usepackage{amssymb}
\usepackage{amsbsy}
\usepackage{amsmath}
\usepackage{stmaryrd}
\usepackage{graphicx}
\usepackage{epsfig}
\usepackage{hyperref}
\makeatletter
\usepackage{pifont}
\makeatother

\begin{document}

\title{Transition from diffusive to ballistic dynamics for a class of finite quantum models}

\author{Robin Steinigeweg}

\email{rsteinig@uos.de}

\affiliation{Fachbereich Physik, Universit\"at Osnabr\"uck,
             Barbarastrasse 7, D-49069 Osnabr\"uck, Germany}

\author{Heinz-Peter Breuer}

\email{breuer@physik.uni-freiburg.de}

\affiliation{Physikalisches Institut, Universit\"at Freiburg,
             Hermann-Herder-Strasse 3, D-79104 Freiburg, Germany}

\author{Jochen Gemmer}

\email{jgemmer@uos.de}

\affiliation{Fachbereich Physik, Universit\"at Osnabr\"uck,
             Barbarastrasse 7, D-49069 Osnabr\"uck, Germany}

\date{\today}

\begin{abstract}
The transport of excitation probabilities amongst weakly coupled
subunits is investigated for a class of finite quantum systems. It
is demonstrated that the dynamical behavior of the transported
quantity depends on the considered length scale, e.~g., the
introduced distinction between diffusive and ballistic transport
appears to be a scale-dependent concept, especially since a
transition from diffusive to ballistic behavior is found in the
limit of small as well as in the limit of large length scales. All
these results are derived by an application of the
time-convolutionless projection operator technique and are
verified by the numerical solution of the full time-dependent
Schr\"odinger equation which is obtained by exact diagonalization
for a range of model parameters.
\end{abstract}


\maketitle

There are certainly many ways in order to analyze the transport
behavior of quantum systems: Linear response theory such as
implemented in the Kubo-formula \cite{kubo1991, mahan1990,
kluemper2002, zotos1997, heidrichmeisner2002, jung2006}, methods
which intend to map the quantum system onto a system of
interacting, classical, gas-like (quasi-)particles
\cite{peierls2001, kohn1957, aoki2006, hornberger2006,
vacchini2005} (possibly represented by Green's functions
\cite{kadanoff1962, rammer1968, baerwinkel1969}), methods which
include reservoirs of the transported quantity, either explicitly
\cite{wang2006} or effectively \cite{saito2003, michel2003,
mejiamonasterio2005}, and probably many more. However, except for
those involving reservoirs, none of these methods allows to
investigate the dependence of transport behavior on the length
scale of a finite system. Unfortunately, the remaining methods
which do involve baths are conceptually more subtle and
computationally rather challenging \cite{saito2003, michel2003,
mejiamonasterio2005}.

In this letter we will discuss a class of finite models which
indeed exhibits scale-dependent transport of, e.~g., excitation
probabilities, energy or particles, according to the respective
interpretations of the model. In particular, we will develop a
method which allows for the complete characterization of the
available types of transport and their dependence on the
considered length scale. It will be especially demonstrated that
the numerically exact solution of the full time-dependent
Schr\"odinger equation perfectly agrees with the theoretical
predictions which are obtained from this method: For intermediate
length scales diffusive behavior is found, that is, the dynamics
of the transported quantity is well described by a respective
diffusion equation and the spatial variance of an initial
excitation profile increases linearly in time. Contrary, in the
limit of small as well as large length scales diffusive behavior
breaks down. This non-diffusive transport type turns out to be
``ballistic'' which is to be understood in terms of a quadratic
increase of the spatial variance. (See the last page of this letter for a
classification of these findings in the framework of standard
solid state theory.)

\begin{figure}[htb]
\includegraphics[width=0.75\linewidth]{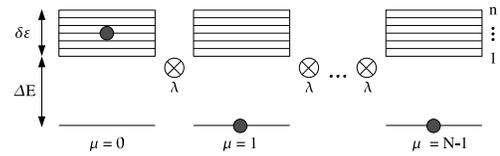}
\caption{A model of $N$ identical, weakly coupled, subunits
featuring a non-degenerate ground state, an energy gap $\Delta E$
and an energy band $\delta \epsilon$ with $n$ equidistant states.
The dots indicate excitation probabilities and are supposed to
visualize a state from the investigated ``single-excitation
space''.} \label{system}
\end{figure}

According to Fig.~\ref{system}, our model consists of $N$
identical subunits which are labelled by $\mu = 0, 1, \ldots,
N-1$. These subunits are assumed to have a non-degenerate ground
state with energy $\varepsilon_0 = 0$ as well as $n$ excited
states with equidistant energies $\varepsilon_i$, $i = 1, 2,
\ldots, n$. We will focus on the (invariant) ``one-excitation subspace'' which
is spanned by the set $\{|\mu,i\rangle\}$, where $|\mu,i\rangle$
corresponds to the excitation ``sitting'' in the $i$th exited
state of the $\mu$th subunit. Using this notation, the
total Hamiltonian $H = H_0 + \, \lambda V$, the sum of a local part $H_0$ and an
interaction part $V$, is given by
\begin{eqnarray}
H_0 &=& \sum_{\mu} \sum_{i} \epsilon_i \, |\mu, i \rangle \langle \mu ,i | \; , \\
V &=& \sum_{\mu} \sum_{i,j} c_{i j} \, | \mu, i \rangle \langle
\mu+1, i| \; + \; \text{H.c.} \label{inter}
\end{eqnarray}
We use periodic boundary conditions, i.~e., we identify $\mu = N$
with $\mu = 0$. The $c_{i j}$ form a normalized Gaussian random
matrix with zero mean, $\lambda$ is an overall coupling constant.
Obviously, $V$ corresponds to nearest-neighbor hopping.

This system may be viewed as a simplified model for, e.~g., a
chain of coupled molecules or quantum dots, etc. In this case the
hopping of an excitation from one subunit to another corresponds
to transport of energy, especially if $\Delta E \gg
\delta\epsilon$. The model may also be viewed as a tight-binding
model for particles on a lattice. In this case the hopping
corresponds to transport of particles. There are $n$ ``orbitals''
per lattice site but no particle-particle interaction in the sense
of a Hubbard model [cf.~Eq.~(\ref{inter})]. Due to the
independence of $c_{i j}$ from $\mu$, these are systems without
disorder in the sense of, say, Anderson \cite{anderson1958}, in
spite of the fact that the $c_{i j}$ are random. For some
literature on this model class, see \cite{michel2005, gemmer2006,
steinigeweg2006, breuer2006}.

The total state of the system is naturally represented by a
time-dependent density matrix $\rho(t)$. We denote by $\Pi_\mu
\equiv \sum_i | \mu, i \rangle \langle \mu ,i|$ the operator which
projects onto the exited states $| \mu ,i \rangle$ of the $\mu$th
subunit's band. Thus, the quantity $P_{\mu}(t) = {\mathrm{Tr}}[ \,
\Pi_\mu \, \rho(t) ]$ represents the probability for finding an
excitation of the $\mu$th subunit, while all other subunits are in
their ground state.

Our aim is to describe the dynamical behavior of these local
probabilities and to develop explicit criteria that enable a clear
distinction between diffusive and ballistic transport. We will
call the behavior diffusive, if the $P_\mu(t)$ fulfill a discrete
diffusion equation
\begin{equation} \label{discdiv}
\dot{P_{\mu}} = \kappa \, (P_{\mu+1} + P_{\mu-1}- 2 \, P_{\mu})
\end{equation}
with some diffusion constant $\kappa$. As well known from the work
of Fourier, such a diffusion equation decouples with respect to,
e.~g., cosine-shaped spatial density profiles, i.~e., the above
equation yields
\begin{equation} \label{divmod}
\dot{F_q} = -2 \, (1 - \cos q) \, \kappa \, F_q \; , \quad F_q
\equiv C_q \sum_{\mu} \cos (q \, \mu) \, P_{\mu}
\end{equation}
with $q = 2 \pi \, k / \, N$, $k = 0, 1, \ldots, N / \, 2$ and an
appropriate normalization constant $C_q$. Hence, if the model
indeed shows diffusive behavior, then all modes $F_q$ have to
relax exponentially. But if the modes $F_q$ are found to relax
exponentially only for some regime of $q$, the model is said to
behave diffusively on the corresponding length scale $l = 2 \pi /
\, q$. The dynamics of the $F_q$, as resulting from the quantum
system, is most conveniently expressed in terms of expectation
values of ``mode operators'' $\Phi_q$,
\\
\begin{equation} \label{F-Q}
F_q(t) = \text{Tr} [ \, \Phi_q \, \rho(t) ] \; , \quad \Phi_q
\equiv C_q \sum_{\mu} \cos (q \, \mu) \; \Pi_\mu \, ,
\end{equation}
where we now choose $C_q = \sqrt{2 / n N}$ for $q \neq 0, \, \pi$
as well as $C_0 = C_{\pi} = \sqrt{1 / n N}$ such that $\text{Tr}[
\, \Phi_q \Phi_{q'} ] = \delta_{q q'}$, the mode operators are
orthonormal.

A strategy for the analysis of the dynamical behavior of the $F_q$
is provided by the projection operator techniques
\cite{kubo1991,nakajima1958,zwanzig1960,breuer2007}. In order to
apply these techniques one first defines an appropriate projection
superoperator ${\mathcal{P}}$. Formally, this is a linear map
which projects any density matrix $\rho(t)$ to a matrix
${\mathcal{P}} \, \rho(t)$ that is determined by a certain set of
relevant variables. Moreover, the map has to be a projection in
the sense of ${\mathcal{P}}^2 = {\mathcal{P}}$. In the present
case a suitable projection superoperator is defined by
\begin{equation}
{\mathcal{P}} \, \rho(t) \equiv \sum_q \text{Tr} [ \, \Phi_q \,
\rho(t) ] \, \Phi_q = \sum_q F_q(t) \, \Phi_q \; .
\end{equation}
Due to the orthonormality of the mode operators this map is indeed
a projection operator. The great advantage of this approach is
given by the fact that it directly yields an equation of motion
for the relevant variables, in our case the Fourier amplitudes
$F_q$. We concentrate here on a special variant of these
techniques which is known as time-convolutionless (TCL) projection
operator method \cite{shibata1979,breuer2007}. Considering initial
states with ${\cal P} \, \rho(0) = \rho(0)$, this method leads to
a time-local differential equation of the form $\dot{F}_q(t) =
-\Gamma_q(t) \, F_q(t)$. So far, this formulation is exact. Modes
with different $q$ do not couple due to the translational
invariance of the model.

Nevertheless, the exact $F_q(t)$ is hard to determine but the TCL
method also yields a systematic perturbation expansion for the
rate $\Gamma_q(t)$ in powers of the coupling constant $\lambda$,
namely, $\Gamma_q(t)=\lambda^2 \, \Gamma_{q,2}(t) + \lambda^4 \,
\Gamma_{4,q}(t) + \ldots$ (all odd order contributions vanish).
Note that a truncation of the above TCL expansion may yield
reasonable results even and especially in the regime where the
elements of the perturbation matrix are larger than the level
spacing of $H_0$, i.~e., the regime where standard
time-independent perturbation theory breaks down.

\begin{figure}[htb]
\includegraphics[width=0.6\linewidth]{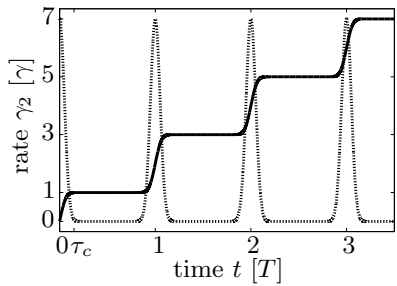}
\caption{Sketch of the correlation function $f(\tau)$ [dashed
line] and its integral $\gamma_2(t)$ [continuous line]: $f(\tau)$
features complete revivals at multiples of $T$ such that
$\gamma_2(t)$ has a step-like form. This sketch indeed reflects
the numerical results for $f(\tau)$ and $\gamma_2(t)$ but
highlights the relevant time scales for clearness. \label{corrfu}}
\end{figure}

In leading order a straightforward calculation yields
\begin{equation} \label{TCL-EQ}
\dot{F}_q(t) = -2 \, (1-\cos q) \, \gamma_2(t) \, F_q(t) \; ,
\end{equation}
where the rate $\gamma_2(t)$ is defined by $\gamma_2(t) = \int_0^t
d\tau \, f(\tau)$ and the two-point correlation function $f(\tau)$
is given by (time arguments refer to the interaction picture)
\[
 f(\tau) = \frac{1}{n} \, \text{Tr} [ \, V(t) \, V(t_1) \, \Pi_\mu ]
 \; , \quad \tau \equiv t - t_1 \; .
\]
This function is completely independent from $\mu$ because of the
translational invariance of the model. Of course, $f(\tau)$
depends on the concrete realization of the random interaction $V$.
But due to the law of large numbers, the crucial features of $V$
are nevertheless the same for the overwhelming majority of all
realizations of $V$, as long as $\sqrt{n} \gg 1$. And in fact,
$f(\tau)$ typically assumes the form in Fig.~\ref{corrfu}. It
decays like a standard correlation function on a time scale of the
order $\tau_c = 1 / \, \delta \varepsilon$. The area under this
first peak is approximately given by $\gamma = 2 \pi \, n \,
\lambda^2 / \, \delta \varepsilon$. However, unlike standard
correlation functions, due to the equidistant level spacing of the
local bands, $f(\tau)$ is a strictly periodic function with the
period $T = 2 \pi \, n / \, \delta \varepsilon$. Consequently, its
time-integral $\gamma_2(t)$ nearly represents a step function, see
Fig.~\ref{corrfu}.

\begin{figure}[htb]
\includegraphics[width=0.6\linewidth]{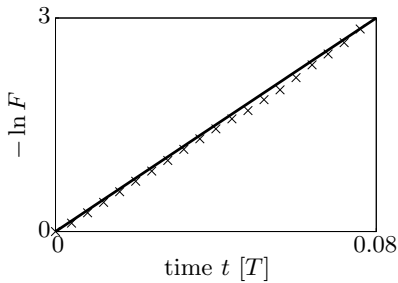}
\caption{Evolution of a Fourier mode $F_{\pi}(t)$ which decays on
an intermediate time scale $\tau_c < t < T$. Numerics (crosses)
shows an exponential decay which indicates diffusive behavior and
is in accord with the theoretical prediction (continuous line).
Parameters: $N = 120$, $n = 500$, $\delta \epsilon = 0.5$,
$\lambda =0.0005$. \label{exdec}}
\end{figure}

Thus, for $\tau_c < t < T$ we find from Eq.~(\ref{TCL-EQ})
\begin{equation} \label{TCL-EQD}
\dot{F}_q(t) = -2 \, (1 - \cos q) \, \gamma \, F_q(t) \; .
\end{equation}
The comparison with Eq.~(\ref{divmod}) clearly indicates diffusive
behavior with a diffusion constant $\kappa = \gamma$. And indeed,
for modes which decay on a time scale $t$ with $\tau_c < t < T$ we
find an excellent agreement between Eq.~(\ref{TCL-EQD}) and the
numerical solution of the full time-dependent Schr\"odinger
equation, obtained by incorporating Bloch's theorem and exactly diagonalizing
the Hamiltonian within decoupled subspaces. A ``typical'' example for a single
realization of the random matrix $V$ is shown in Fig.~\ref{exdec}. Note
that the size of the system is chosen adequately large such that
there are only small fluctuations of the order of $1/n$ which
arise due to the discrete spectrum, see, e.~g.,
\cite{michel2005, gemmer2006}.

\begin{figure}[htb]
\includegraphics[width=0.75\linewidth]{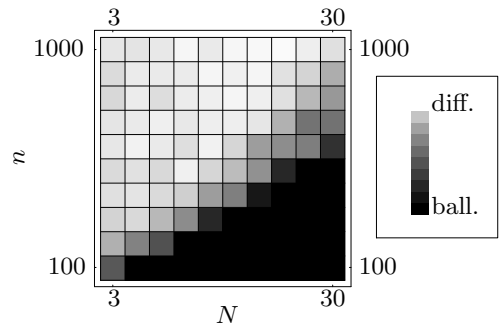}
\caption{Deviations of the time evolution of the Fourier modes
$F_q(t)$ with $q = 2 \pi/N$, the longest wavelength, from a purely
exponential decay for different model parameters $N$ and $n$.
These deviations are based on a measure used in
\cite{steinigeweg2006} and are in accord with the claim that
diffusive transport behavior is restricted to the regime defined
by condition (\ref{cond1}). Other model parameters: $\delta
\epsilon = 0.5$, $\lambda = 0.0005$.} \label{phadi}
\end{figure}

However, until now the above picture is not complete for two
reasons. The first reason is that $F_q(t)$ may decay on a time
scale that is long compared to $T$. According to
Eq.~(\ref{TCL-EQ}), this will happen, if $2(1-\cos q)\gamma T \gg
1$ is violated. If we approximate $2(1 - \cos q) \approx q^2 = 4
\pi^2/l^2$ for rather small $q$ (large $l$), this leads to the
condition
\begin{equation} \label{cond1}
\left( \frac{4 \pi^2 \, n \, \lambda}{l \, \delta \varepsilon}
\right)^2 \gg 1 \; .
\end{equation}
If this condition is satisfied for the largest possible $l$,
i.~e., for $l=N$, the system exhibits diffusive behavior for all
modes. If, however, the system is large enough to allow for some
$l$ that violates condition (\ref{cond1}), diffusive behavior
breaks down in the long-wavelength limit. This result is again
backed up by numerics, see Fig.~\ref{phadi}.

\begin{figure}[htb]
\includegraphics[width=0.6\linewidth]{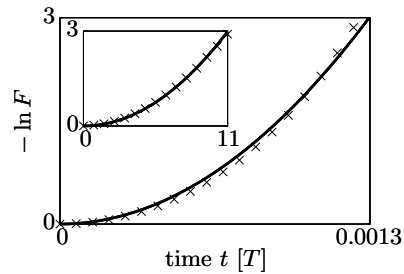}
\caption{ Evolution of a Fourier mode $F_{\pi}(t)$ which decays on
a time scale $t \ll \tau_c$, parameters: $N =120$, $n=500$, $\delta
\epsilon=0.5$, $\lambda=0.004$. Inset: Evolution of a Fourier mode
$F_{\pi/60}(t)$ which decays on a time scale $t \gg T$, parameters:
$N =120$, $n=500$, $\delta \epsilon=0.5$, $\lambda=0.0005$. In both
cases numerics (crosses) shows a Gaussian decay which indicates ballistic
behavior and is in accord with the theoretical predictions (continuous
lines).\label{gauss}}
\end{figure}

Towards what transport type does the system deviate from
diffusive, if condition (\ref{cond1}) is violated? In this regime
we have to consider time scales with $t \gg T$, as already
mentioned above. We may thus approximate $\gamma_2(t)\approx
2\gamma t/T$, see Fig.~\ref{corrfu}. Plugging $\kappa = 2\gamma
t/T$ into Eq.~(\ref{discdiv}) yields a spatial variance $\sigma^2
\equiv \langle \mu^2 \rangle - \langle \mu \rangle^2= 2\gamma
t^2/T$, contrary to $\sigma^2 = 2\gamma t$ which results for
$\kappa = \gamma$. ($\kappa =\gamma$ applies in the regime where
condition (\ref{cond1}) is satisfied.) This change of the
time-scaling of $\sigma^2$ clearly indicates the transition from
diffusive to ballistic transport. The validity of the TCL approach
in the ballistic regime is again backed up by numerics: Here, the
TCL theory obviously predicts a Gaussian decay of $F_q(t)$ which
is in agreement with the numerical results, see Fig.~\ref{gauss}.

In a second case transport may be non-diffusive, if the
corresponding $F_q(t)$ decay on a time scale that is short
compared to $\tau_c$. This will happen, if $2 \, (1-\cos q) \,
\gamma \, \tau_c \ll 1$ is violated. We may approximate $2 \, (1-
\cos q) \approx 4$ for the largest possible $q$ (smallest possible
$l$). Hence, the above inequality may be written as
\begin{equation} \label{cond2}
\frac{8 \pi \, n \, \lambda^2}{\delta \varepsilon^2} \ll 1 \; .
\end{equation}
If this inequality is violated, diffusive behavior breaks down in
the limit of short-wavelength modes. Moreover, if the second order
still yields reasonable results for not too large $\lambda$, we
expect a linearly increasing rate $\gamma_2(t)$ and thus a
Gaussian decay, that is, according to the above reasoning,
ballistic transport. For increasing wavelength, however, the
corresponding inequality will eventually be satisfied, thus
allowing for diffusive behavior. Also these conclusions are in
accord with numerics, see Fig.~\ref{gauss}.

Standard solid state theory always predicts ballistic transport
for a translational invariant model without particle-particle
interactions. Nevertheless, in the limit of many bands (many
orbitals per site) and few sites (few $k$-values) two features may
occur: i) The band structure in $k$-space becomes a disconnected
set of points rather than the usual set of distinct smooth lines.
It is hence impossible to extract velocities by taking the
derivatives of the dispersion relations. ii) The eigenstates of
the current operator no longer coincide with the Bloch eigenstates
of the Hamiltonian and the current becomes a non-conserved
quantity, even without impurity scattering. It is straightforward
to check that both features occur in the regime where condition
(\ref{cond1}) is fulfilled. This is the regime where standard
solid state theory breaks down due to the fact that the system is
to ``small''.

On the other hand, our numerical simulations clearly reveal that
the TCL projection operator technique is applicable both in the
diffusive and in the ballistic regime and that it correctly
describes the transition between these regimes. Recall that all
analytical results have been obtained from the second order
contribution [Eq.~(\ref{TCL-EQ})] of the TCL expansion. And in
fact, it is possible to demonstrate that higher-order
contributions are confidently negligible. We omit the details of
the determination of the higher orders of the TCL expansion
(several examples are discussed in \cite{breuer2007}), since this
is beyond the scope of this letter and since the excellent
agreement with the numerical simulations is surely evident.

However, the fact that already the second-order TCL equation
(\ref{TCL-EQ}) yields an excellent quantitative agreement with the
numerics, by contrast to standard perturbation theory, can be
understood from the following argument. We first note that for
$\lambda=0$ the energy eigenvalues of the Hamiltonian are $N$-fold
degenerate, corresponding to the resonant transitions
$|\mu,i\rangle \rightarrow |\mu\pm 1,i\rangle$ between neighboring
subunits induced by the interaction. While standard perturbation
theory is spoiled by the presence of these resonant transitions,
the TCL expansion works perfectly well even in the case of exact
degeneracies. In fact, an analysis of the structure of the
higher-order correlation functions reveals that for an interaction
Hamiltonian $V_d$ which includes only the resonant transitions,
the second-order TCL equation (\ref{TCL-EQ}) is {\textit{exact}}
on average (for arbitrary $\lambda$) with small fluctuations of
order $1/n$. This is due to the fact that the TCL method is based
on an expansion in terms of the ordered cumulants of $V_d$
\cite{breuer2007}, the higher-orders of which vanish by virtue of
the Gaussian character of the coupling matrix elements. This
implies that the second order TCL equation already reproduces
exactly the effect of all resonant transitions, which is the
crucial reason for the success of the TCL projection operator
approach.

\begin{acknowledgments}
We sincerely thank H.-J.~Schmidt and A. Rosch for fruitful
discussions. Financial support by the Deutsche
Forschungsgemeinschaft is gratefully acknowledged.
\end{acknowledgments}

\end{document}